\documentclass[5p]{elsarticle} 

\usepackage{mathtools}
\usepackage[table,xcdraw]{xcolor}
\journal{Energy}

\usepackage{placeins}

\usepackage{eurosym}
\usepackage{threeparttable} 
\usepackage{subcaption}
\usepackage{url}
\usepackage{xurl} 
\usepackage[colorlinks=true, citecolor=blue, linkcolor=blue, filecolor=blue,urlcolor=blue]{hyperref}
\usepackage[acronym, automake, nonumberlist]{glossaries-extra}

\usepackage{lineno}
\modulolinenumbers[5]

\usepackage{lineno,hyperref}
\modulolinenumbers[1]
\usepackage{amsmath}
\usepackage[T1]{fontenc}
\usepackage{siunitx}
\usepackage{eurosym}
\biboptions{numbers,sort&compress}
\usepackage[europeanresistors,americaninductors]{circuitikz}
\usepackage{adjustbox}
\usepackage{xspace}
\usepackage{caption}
\usepackage{booktabs}
\usepackage{tabularx}
\usepackage{threeparttable}
\usepackage{multicol, multirow}
\usepackage{float}
\usepackage{graphicx,dblfloatfix}
\usepackage{csvsimple}
\usepackage{amssymb}
\usepackage{pifont}
\usepackage{tikzsymbols}
\usepackage{textcomp}

\newcommand{\cmark}{\ding{51}}%
\newcommand{\xmark}{\ding{55}}%

\newcolumntype{L}{>{\begin{math}}l<{\end{math}}}%
\newcolumntype{C}{>{\begin{math}}c<{\end{math}}}%

\newcommand{\mwth}{\ \officialeuro /MWh$_{\text{th}}$}
\newcommand{\mwh}{\ \officialeuro /MWh\ }

\makeglossaries
\newacronym{tes}{TES}{Thermal energy storage (in form of water tanks)}
\newacronym{chp}{CHP}{Combined heat and power plant}
\newacronym{cop}{COP}{Coefficient of Performance}
\newacronym{bdew}{BDEW}{Bundesverband der Energie- und Wasserwirtschaft}
\newacronym{PV}{PV}{Solar photovoltaics}
\newacronym{ghg}{GHG}{Greenhouse gas}
\newacronym{EU}{EU}{European Union}
\newacronym{dea}{DEA}{Danish Energy Agency}
\newacronym{helmeth}{HELMETH}{Integrated High-Temperature Electrolysis and Methanation for Effective Power to Gas Conversion}
\newacronym{jrc}{JRC}{Joint Research Center}
\newacronym{idees}{IDEES}{Integrated Database of the European Energy System}
\newacronym{ecmwf}{ECMWF}{European Centre for Medium-Range Weather Forecasts}
\newacronym{ocgt}{OCGT}{Open cyclic gas turbine}
\newacronym{v2g}{V2G}{Vehicle to Grid}
\newacronym{pypsa}{PyPSA}{Python for Power System Analysis}
\newacronym{tyndp}{TYNDP}{Ten-year network development plan}
\newacronym{entso-e}{ENTSO-E}{European network for transmission system operators electricity}
\newacronym{nuts}{NUTS}{Nomenclature of Territorial Units for Statistics}
\newacronym{covid}{COVID-19}{Coronavirus disease 2019}
\newacronym{entsog}{ENTSO-G}{European Network of Transmission System Operators for Gas}
\newacronym{dsm}{DSM}{Demand side management}
\newacronym{hvdc}{HVDC}{High voltage direct current}
\newacronym{phs}{PHS}{Pumped hydro storage}
\newacronym{dac}{DAC}{Direct air capture}
\newacronym{sng}{SNG}{Synthetic natural gas}
\newacronym{ise}{Fraunhofer ISE}{Fraunhofer Institute for Solar Energy Systems}
\newacronym{diw}{DIW}{German Institute for Economic Research (Deutsches Institut f\"ur Wirtschaftsforschung)}
\newacronym{SMR}{SMR}{Steam methane reforming}
\newacronym{fom}{FOM}{Fixed operation and maintenance costs}
\newacronym{vom}{VOM}{Variable operation and maintenance costs}
\newacronym{KKT}{KKT}{Karush-Kuhn-Tucker}
\newacronym{hdd}{HDD}{Heating degree days [d/a]}
\newacronym{tabula}{TABULA}{Typology Approach for Building Stock Energy Assessment}
\newacronym{EN}{EN}{European Norm}
\newacronym{ISO}{ISO}{International Organization for Standardization}

\begin{document}

\begin{frontmatter}

\title{Mitigating heat demand peaks in buildings in a highly renewable European energy system}

\author[kitaddress]{Elisabeth Zeyen}
\author[kitaddress]{Veit Hagenmeyer}
\author[tuaddress,kitaddress]{Tom Brown}
\address[kitaddress]{Institute for Automation and Applied Informatics (IAI), Karlsruhe Institute of Technology (KIT), Forschungszentrum 449, 76344, Eggenstein-Leopoldshafen, Germany}
\address[tuaddress]{Department of Energy Systems, Faculty of Process Engineering, Einsteinufer 25 (TA 8), 10587 Berlin, Germany}

\begin{abstract}
Space and water heating accounts for about 40\% of final energy consumption in the European Union and thus plays a key role in reducing overall costs and greenhouse gas emissions. Many scenarios to reach net-zero emissions in buildings rely on electrification, but meeting the heat demand peaks in the winter can be challenging, particularly when wind and solar resources are low. This paper examines how to mitigate space heating demand peaks most cost-effectively in a top-down, sector-coupled model with carbon dioxide emissions constraint to be net-zero.  It introduces the first model that co-optimises both supply and efficiency simultaneously including all European countries with hourly resolution. The competition between technologies to address these heating peaks, namely building retrofitting, thermal energy storage and individual hybrid heat pumps with backup gas boilers is examined. A novel thought experiment demonstrates that the level of building renovation is driven by the strong seasonal heat peaks, rather than the overall energy consumption. If all three instruments are applied, total costs are reduced by up to 17\%. Building renovation enables the largest benefit with cost savings of up to 14\%  and allows individual gas boilers to be removed from the energy system without significant higher costs.
\end{abstract}
\begin{keyword}
building retrofitting; space heating; energy system modelling; sector coupling; optimisation
\end{keyword}
\end{frontmatter}


\section*{Highlights}
\begin{itemize}
	\item energy supply and efficiency measures are co-optimised in a highly-renewable European energy system with net-zero CO$_2$ emissions
	\item cost-optimal solution includes building renovations to save an average of 44-51\% of space heat depending on the country
	\item with efficiency improvements, all gas distribution to buildings can be excluded with only 1-2\% higher costs
	\item investments in energy efficiency in buildings are driven by need to reduce peak demand, not overall energy consumption
	\item up to 30\% of total system costs are caused by seasonal heat demand peaks
\end{itemize}

\section{Introduction}
A major part of the \gls{EU}'s final energy consumption is used to supply heat in buildings. In 2018, 75\% of this energy was delivered by fossil fuels, generating 36\% of the \gls{EU}'s total CO$_2$ emissions \cite{co2_emissions, jrc_taylor}. A structural change of the heating sector is needed not only to avoid environmental impacts caused by emissions, but also to address social challenges such as affordable heating and health of the residents. Furthermore a transition to renewable technologies could mitigate dependencies on fossil fuel imports. The literature on the heating transition is now reviewed, with a focus on important gaps that are addressed by the present paper.\\
\\ Numerous studies have already examined how to transform the heating sector. Most of them identify as key factors large-scale electrification \cite{heat_roadmap3, henning_palzer2, bogdanov_2021} and a stronger coupling between the sectors electricity and heat \cite{iee_frauenhofer}. In an energy system highly based on renewable sources, this transition raises new challenges \cite{papadis2020}. A key issue is the weather dependency of both thermal supply and demand through an increased electrification. On days with cold temperature, calm wind and low solar radiation (so called `cold dark wind lulls'), the space heating demand is high while the feed-in of wind and solar, as well as the efficiency of air-sourced heat pumps, is low. Furthermore, the heating demand of buildings compared to the other sectors electricity, transport and industry, is more seasonal and possesses stronger peaks. Several studies \cite{heat_roadmap3, staffell2018} have determined the increase in heat peak demand. But, despite its importance, none of these studies has quantified the costs and impacts on system design of these heat demand peaks in a sector-coupled European model. The European perspective is important, because energy markets are already strongly coupled internationally, and a cross-sectoral perspective is necessary to understand, for example, the impact of heat pump demand on electricity supply. \\
\\To enable the decarbonization of the heating sector, renovation of buildings is widely considered as a key factor. About 75-90\% \cite{primes_eu_2018} of the current building stock is still going to be used in 2050, and therefore it is necessary to renovate the existing houses. As part of the European Green Deal \cite{green_deal}, the \gls{EU} has called for a renovation wave, setting as goal to at least double the current renovation rate of 1\% of buildings per year to decarbonise the residential and non-residential building sector \cite{reno_wave, ltrs_eu}. Work on the potential of heat demand saving through building renovation has been carried out on a national level for Germany by Henning et al. \cite{henning_palzer2}. They find 50\% savings in space heat demand to be cost-optimal. On a European level, several authors \cite{heat_roadmap2, Broin2013} see 30-50\% of space heat demand reduction. However, the existing studies do not capture the role of renovation at the European level because they are limited to either individual countries, the residential sector or exogenous assumptions about energy supply costs or the extent of building renovation. Given the strong impact that renovations have on the shape of demand curves, such demand-side measures should to be co-optimised with supply options.\\
\\ Hybrid heating, i.e. the use of multiple heating technologies in a building, has been identified as a further tool for balancing peak demands. Clegg et al. \cite{Clegg2018} find for the UK with an assumed 75\% CO$_2$ emissions reduction, that hybrid heating technologies can reduce generation peaks by 24\%. Thermal energy storage (\gls{tes}) is also regarded as an important component to allow the transition of the heating sector \cite{arteconi2013}. Hedegaard et al. \cite{hedegaard2013} analyse the cost and peak shaving benefits of flexible heat pump operation with thermal storage in Denmark. Malley et. al \cite{malley2} show in the case of Ireland that \gls{tes} can bridge the so called `cold dark wind lulls' and enable higher feed-in of wind. Previous work is limited to single countries and does not capture the competition between hybrid heating, thermal storage and building efficiency. In a European model Brown et al. \cite{synergies} and in a global model  Boganov et al. \cite{bogdanov_2021} show that \gls{tes} can be used as a long-term seasonal storage in district heating networks and as a short-term individual storage to smooth peak demand and thus reduce overall costs. However, these studies do not include demand-side efficiency measures to examine how they would compete with supply-side and storage options.\\
\\The present paper directly tackles the gaps in the literature  identified in the previous paragraphs. In contrast to the prior literature, this paper addresses the decarbonisation of building heating, and the particular problem of `cold dark wind lulls', by explicitly modelling the competition between investments in heating electrification, hybrid heating with backup gas boilers, building renovation and thermal storage in a European system with net-zero CO$_2$ emissions. We introduce the first model that co-optimises supply-side options and efficiency measures in buildings with country-level resolution and hourly modelling. This co-optimization is important: models that assume exogenous energy prices miss how the energy supply can adapt to the new demand profile, while models that assume exogenous efficiency measures may arrive at sub-optimal combinations of supply and efficiency measures. The important synergies between the sectors are incorporated by applying a fully sector-coupled model, including electricity, heat for buildings and land transport. Building renovation is modelled top-down and considers both the residential and the often neglected service sector. Costs and energy-savings are determined based on country-specific (and not just climate-zone specific \cite{large_scale_sav, entranze}) characteristics. To identify the systemic importance of seasonal heating peaks, we introduce a novel thought experiment that compares the impacts and costs of these peak demands to a heat profile with no seasonal variations. The robustness of the results is analysed using a sensitivity analysis with regard to different historical weather years and the costs of building renovation. \\
\\These model innovations allow us to address the following important research questions regarding the decarbonisation of the heating sector:
\begin{itemize}
	\item How do building retrofitting, thermal energy storage (\gls{tes}) and individual hybrid heat pumps with backup gas boilers compete to meet heat demand peaks?
	\item What effect does this competition have on carrier composition and costs?
	\item To what extent do heat demand peaks influence the costs and design of a highly-renewable European energy system?
\end{itemize}
By answering these questions in a co-optimised cross-sectoral European model, we can for the first time capture important interactions and synergies between demand and supply, between sectors, and in international energy trading.
\section{Methods}
In the first part of the Methods section (\ref{sec:model}), we introduce the overarching sector coupling model PyPSA-Eur-Sec. Secondly, we describe in more detail how we model demand (\ref{sec:method_heating}) and supply (\ref{sec:heat_supply}) in the heat sector, which is the focus subject of our analysis. In the final part (\ref{sec:retro}), we outline how the country-specific costs for space heat demand savings are determined. 
\subsection{Model description}\label{sec:model}
\begin{figure}[h]
	\centering
	\includegraphics[width=1.0\linewidth]{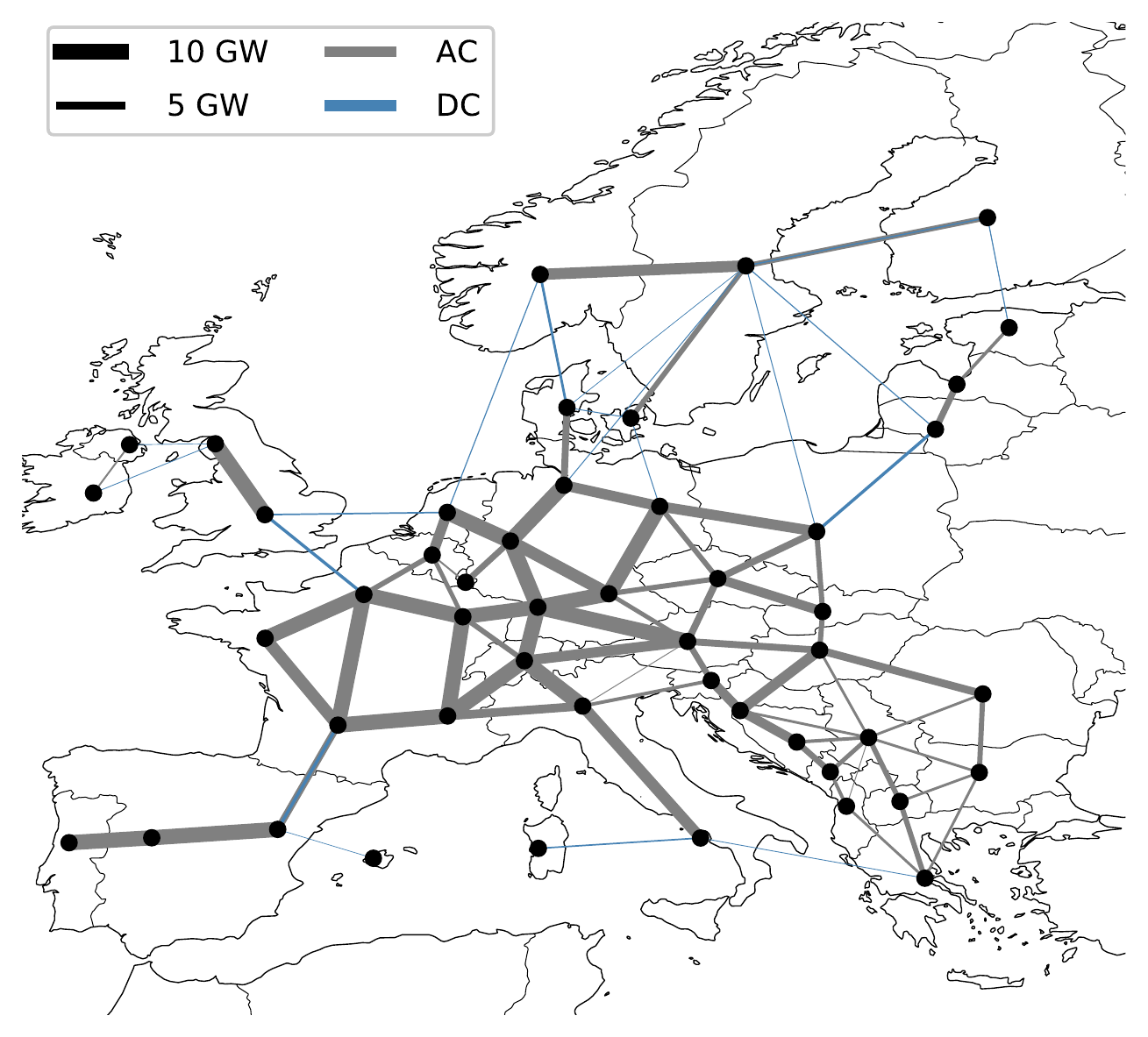}
	\caption{The European 48-node model with the electricity transmission network, based on existing network capacities (see section \ref{sec:model}).}
	\label{fig:topology}
\end{figure}

We use the least-cost capacity expansion model \gls{pypsa}-Eur-Sec presented in \cite{synergies} which represents a comprehensive sector coupling approach (electricity, heating and land transport) including high resolution in space and time. 
The model is constructed to represent a net-zero CO$_2$ emissions target, corresponding to European policy, for the year 2050. It is based on the free framework Python for Power System Analysis (\gls{pypsa}) \cite{pypsa} and the open optimisation model for the European transmission system \gls{pypsa}-Eur \cite{PyPSAEur}. It is important to cover the whole of Europe, since the supply of energy is strongly affected by energy trading between countries. The model has at least one node for each country in the interconnected European electricity system. Large countries like Germany and France have more nodes to represent country-internal grid bottlenecks. For the simulations we chose a total of 48 nodes for Europe (see figure \ref{fig:topology}) as a compromise between representing grid bottlenecks while maintaining computational tractability.
 Total costs are minimised applying linear optimisation to reach cost-efficient solutions while maintaining physical, technical and socio-economic constraints. It includes electricity generation (rooftop and utility solar photovoltaic (\gls{PV}), onshore wind, offshore wind, methane), storage (battery, hydrogen, methane, thermal storage in hot water tanks, carbon dioxide), heating in buildings (heat pumps, resistive heater, combined heat and power plants (\gls{chp}), solar thermal collectors, building retrofitting), energy converters  (electrolysis, methanation, steam methane reforming (\gls{SMR})) and grid (electricity transmission and distribution, hydrogen, methane, biomass transport). These are all subject to siting and capacity optimisation. CO$_2$ emissions are limited to be net-zero. All the considered technologies with their corresponding costs and conversion efficiencies are taken where possible from the Danish Energy Agency (\gls{dea}) \cite{cost_dea}.\\
%
\begin{figure}[H]
	\centering
	\includegraphics[width=1.0\linewidth]{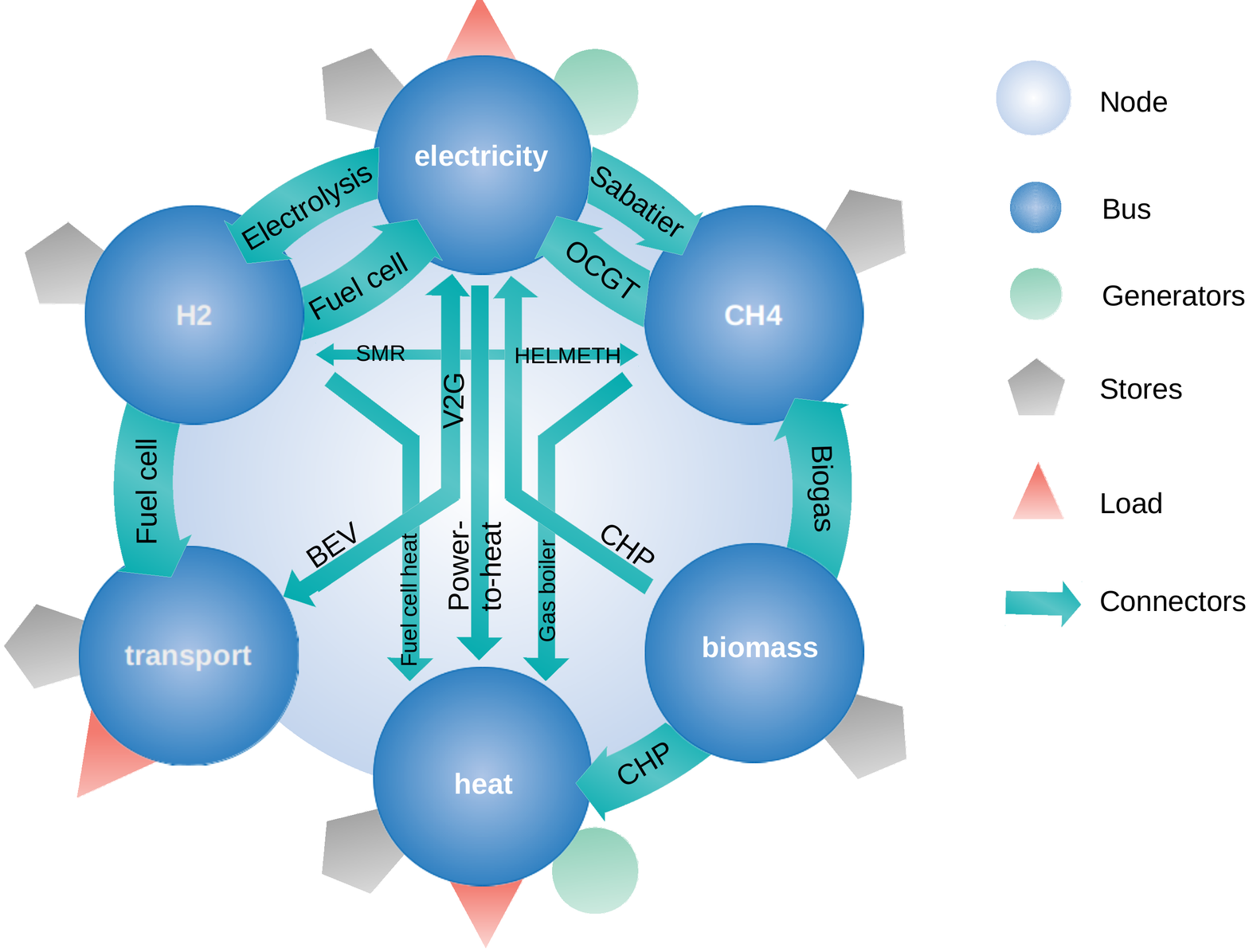}
	\caption{Overview of the sector-coupled model \gls{pypsa}-Eur-Sec. Each region is represented by a node which in turn consists of different `buses' that represent different energy carriers (e.g. hydrogen). Energy can be converted between the buses (e.g. by hydrogen electrolysis). Further details are given in section \ref{sec:model}.}
	\label{fig:model}
\end{figure}
 Each of the 48 nodes consists of individual `buses'. These `buses' represent the different energy carriers, namely electricity (separated into transmission and distribution), heat (divided into one bus for district and four buses for individual heat networks split by sector and population density), transport, hydrogen, methane and biomass (see figure \ref{fig:model}). Energy can be transferred within one node through connectors (for example, electricity can be converted into hydrogen through electrolysis). Energy (transport) networks for electricity, hydrogen, and methane transmission as well as biomass land transport connect buses of the respective carrier between nodes. \\
\\Over one year of hourly operation, total annual costs are minimised. The total costs consists of fixed and operational costs. Fixed annualised investment costs $c_{n,s}$ of the energy capacities $E_{n,s}$, power capacities $G_{n,s}$ and $c_{l}$ inter-connector capacities $F_{l}$ are considered. Operational costs $o_t$ are defined for power dispatch through generation and storage $g_t$, as wells as power flow through inter-connectors $f_t$ at every time step $t$. The objective function with buses $n$, generation and storage technologies $s$, bus connectors $l$ is defined as
\begin{align}\label{eq:objective}
\min_{G,E,F,g,f} \bigg[ &\sum_{n,s} c_{n,s}\cdot G_{n,s} + \sum_{n,s}  \widehat{c}_{n,s}\cdot E_{n,s} + \sum_{l}  c_{l}\cdot F_{l} + \\ 
&\sum_{n,s,t} o_{n,s,t} \cdot g_{n,s,t} + \sum_{l, t} o_{l,t} \cdot f_{l,t}  \bigg]. \nonumber
\end{align}
The energy demand $d_{n,t}$ at each bus $n$ must be met at each time $t$ by either local generators and storage or by a flow $f_{l,t}$ from the connectors $l$
\begin{equation}\label{eq:KKT}
\sum_s g_{n,s,t} + \sum_l a_{l,n,t} \cdot f_{l,t} = d_{n,t} \quad \leftrightarrow \lambda_{n,t} \quad \forall n,t ,
\end{equation}
where $a_{l,n,t}$ represents the flow direction and connector efficiency $\eta_{l,t}$. It is $a_{l,n,t}=-1$ if $l$ starts at $n$ and $a_{l,n,t}=\eta_{l,t}$ if $l$ ends at $n$. The Karush-Kuhn-Tucker (\gls{KKT}) multiplier $\lambda$ represents the market price of the energy carrier at bus $n$ and hour $t$. Electricity and transport demand are assumed to be completely inelastic. Heat demand can be reduced by retrofitting the thermal envelope of buildings. Renovation measures are modelled like a supply technology with heat output $g_{n,\text{retro}, t}$ which reflects a constant fraction $r_n \in [0, r_{n, max}]$ of the time-varying space heat demand $d_{n,t}$ at the bus $n$
\begin{equation}\label{eq:retro_gen}
g_{n,\text{retro}, t} = r_n \cdot d_{n,t}.
\end{equation}
For instance, if in the cost-optimal case the space heat demand $d_{n,t}$ is reduced by 20\% through renovation measures, $r_n=0.2$ and $g_{n,\text{retro}, t}$ represents the space heat saved per hour.
The costs for retrofitting the thermal envelope are included in the objective function (equation \ref{eq:objective}) as the efficiency measures are modelled as generators with fixed costs per heat saving. The assumed costs for retrofitting the thermal envelope with corresponding maximum energy savings $r_{n, max}$ which depend on building stock and sector, are further described in section \ref{sec:retro}.\\
 \\The objective function (equation \ref{eq:objective}) is subject to the following additional constraints which are introduced in detail in previous papers \cite{PyPSAEur, synergies}: (i) technical and geographical limitations on maximum capacities for generators, storage and transmission, (ii) the availability of renewable generation at each bus and time based on weather data, (iii) storage consistency equations, (iv) constraints on the active linearised power flow and (v) net-zero carbon dioxide emissions. \\
\\ The model has been comprehensively introduced in \cite{synergies} and further information are given in the supplementary material. In the following only the newly developed features for this study are addressed.
 \subsubsection{Heat demand} \mbox{}\label{sec:method_heating} \\
 Low-temperature space and water heating is considered, distinguished into two sectors: residential and services. The industrial sector is not included and the required energy for cooking is incorporated in the electricity demand. Total annual load per country is based on data provided by `Integrated Database of the European Energy System' (\gls{jrc}-\gls{idees}) \cite{jrc-idees}, Eurostat \cite{eurostat_energybalances} and additional statistics for Switzerland \cite{swiss_stats} and Norway \cite{norway_stats}. Regarding countries with missing data for space and water heating, the average share based on the other countries is assumed. The overall heat demand per country is distributed spatially weighted by population density and temporally by creating time-series based on ambient air temperature and typical end-user behaviour. \\
 \\ \textbf{Spatial disaggregation}: Heat demand is spatially distributed within a country weighted by the population density. At each node a distinction is made between rural and urban areas. Such a distinction is chosen since different heating systems are present in the corresponding region (e.g. \gls{chp}s in urban district heating networks). This is done by calculating the population density at \gls{nuts}3 level. For each country, starting with the \gls{nuts}3 region with the lowest population density, all regions are counted as rural until the country-specific proportion of the population living in rural areas is reached. The remaining \gls{nuts}3 regions with higher population density are counted as urban. The \gls{nuts}3 regions are then assigned to the 48 nodes and the ratio of the population in rural and urban areas per bus is calculated. Urban areas are further subdivided into regions with individual and district heating, according to today's district heating share  \cite{euroheat} of the corresponding country. Losses in the district heating system are considered to be 15\% \cite{dh_losses}. \\
 \\ \textbf{Temporal disaggregation}: The time series of space heat demand are determined in two steps. First, diurnal profiles are calculated by applying the degree day approximation to every region using daily averaged ambient air temperature data from \gls{ecmwf} ERA5 \cite{era5} and a  threshold $T^{\text{thres}}$=$15^o$ Celsius. Second, to reflect typical end-user behaviour, hourly patterns from `Bundesverband der Energie- und Wasserwirtschaft' (\gls{bdew}) \cite{heat_profile} are used which vary by sector (residential or service) and by weekday (working day or weekend). The resulting hourly resolved load profile is scaled to the total space heating demand of the respective region. Hot water demand is assumed to be constant during the year.

 \subsubsection{Heat supply}\label{sec:heat_supply} 
\begin{figure}
 	\centering
 	\includegraphics[width=1.0\linewidth]{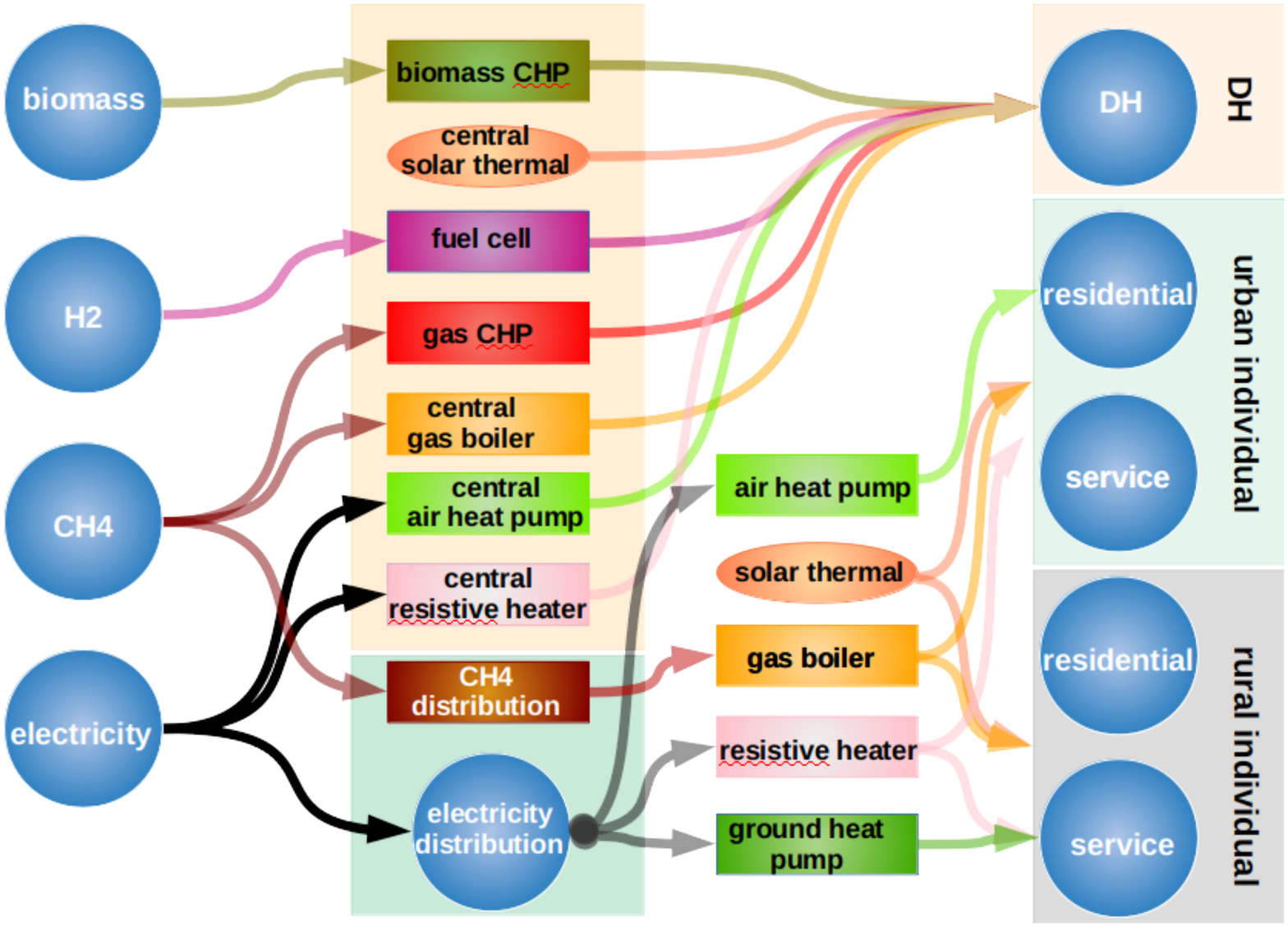}
 	\caption{Heat can be supplied either biomass, solar thermal collectors, gas (hydrogen, methane)  or power-to-heat. At each node, the demand is split by high and low population density (urban and rural). Urban areas are further divided into district heating (\textit{DH}) and individual heating. Further details are given in section \ref{sec:heat_supply}.}
 	\label{fig:heat_supply}
 \end{figure}

\mbox{} \\
 Thermal demand can be met by the following four types: (i) power-to-heat (heat pumps, resistive heater), (ii) gas-to-heat (\gls{chp}, gas boiler, fuel cell) using synthetic gas, upgraded biogas or hydrogen, (iii) solar thermal collectors and (iv) solid biomass-to-heat in district heating networks (\gls{chp}s) (see figure \ref{fig:heat_supply}). \\
 %
 \\ Heat pumps of two categories are considered: ground-sourced brine-to-water heat pumps, which are restricted to rural regions due to land requirements and air-to-water heat pumps usable in areas with high population density. Heat pumps have a poorer efficiency with lower ambient temperatures. This relationship is of particular relevance (since the heat demand rises with colder weather) and can be expressed by the coefficient of performance (\gls{cop}). The  \gls{cop} time-series are calculated by relations from \cite{cop}. They depend on ambient and sink water temperature, which is set to $T_{\text{sink}}=55^o$C \cite{sink_t_pump}.\\
 \\ \gls{chp} powered by synthetic gas, upgraded biogas or solid biomass, as well as fuel cells operating with hydrogen can only supply heat in regions with district heating. Heating via gas in decentralised systems can be provided by either gas boilers or micro \gls{chp}s. Fossil fuels and imports are not considered in this study. Therefore synthetic gas and hydrogen have to be obtained by methanation or electrolysis within Europe. Two different methanation processes are modelled: the Sabatier and the `Integrated High-Temperature Electrolysis and Methanation for Effective Power to Gas Conversion' (\gls{helmeth}) process \cite{helmeth}. In the Sabatier process, hydrogen (H$_2$) and carbon dioxide (CO$_2$) react exothermically to methane (CH$_4$) and water (H$_2$O). The used hydrogen must first be produced by electrolysis. The \gls{helmeth} process combines these two steps by using the heat of the exothermic methanation reaction in a steam electrolysis to generate hydrogen. \\
 \\Thermal energy storage (\gls{tes}) in the form of water tanks is available as a seasonal storage ($\tau=180$ days) in urban areas with district heating, while in decentralised heated regions there is only short-therm storage ($\tau=3$ days). $\tau$ represents the time constant of the decay of thermal energy. The energy lost per hour is $1-\exp(-\frac{1}{24\tau})$.
\subsubsection{Retrofitting of the thermal envelope of buildings}\label{sec:retro} 
\begin{figure}
	\centering
	\includegraphics[width=1.0\linewidth]{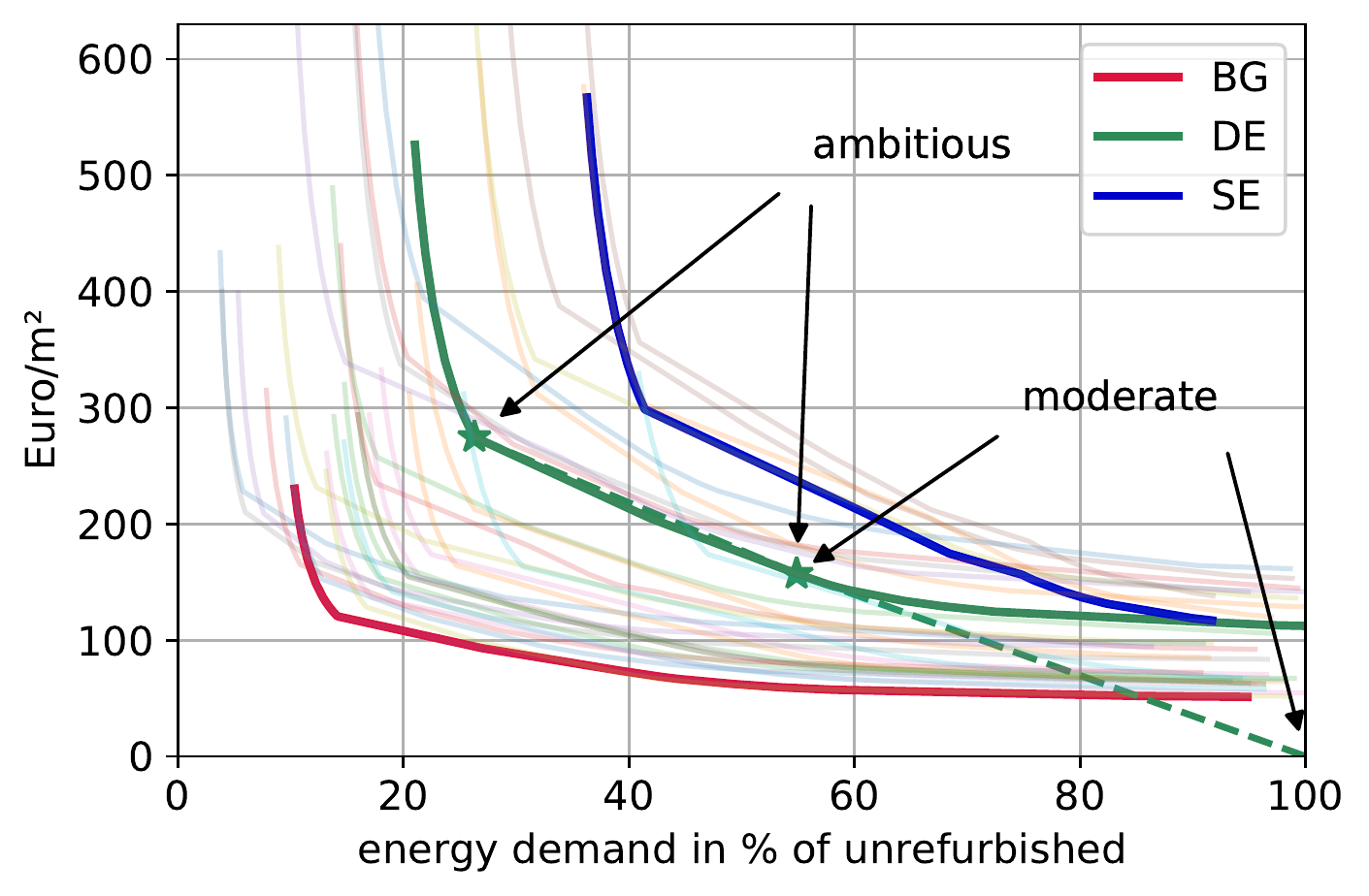}
	\caption{Overnight costs for retrofitting of the thermal envelope of buildings per m$^2$ for energy savings including the residential and the service sector. The costs of all investigated countries are shown in the background to illustrate the scope. Especially highlighted are the curves of Sweden (SE), Germany (DE) and Bulgaria (BG), with the highest costs for retrofitting in Sweden, followed by Germany and Bulgaria. In the optimisation these curves are linearised in two steps (e.g. hashed line for Germany), see section \ref{sec:retro}.}
	\label{fig:energy_cost_curve}
\end{figure}

\mbox{} \\
Renovation of the thermal envelope reduces the space heating demand and is optimised at each node for every heat bus. We consider renovation measures through additional insulation material and replacement of energy inefficient windows. In a first step, costs per energy savings are estimated. They depend on the insulation condition of the building stock and costs for renovation of the building elements. In a second step, for those cost per energy savings two possible renovation strengths are determined: a moderate renovation with lower costs and lower space heat savings, and an ambitious renovation with associated higher costs and higher efficiency gains. They are implemented into the sector coupling model through stepwise linearisation (see figure \ref{fig:energy_cost_curve}). In this study, heat supply technologies, thermal storage and prices for energy supply are optimised and not part of the energy savings cost functions. \\
\\ First, costs per energy savings are calculated based on country-specific characteristics of the building stock, taking into account age and building types, current energy efficiencies and heated floor areas \cite{hotmaps, EU_building}. The buildings are categorized in the following groups:
\begin{itemize}
	\item 9 building categories:\\
	3 residential:
	single family house, multi family house, apartment blocks \\
	6 service:
	offices, trade, education, health, hotels and restaurants, other service buildings
	\item 7 construction periods: \\
	before 1945, 1945-1969 and then in ten-years intervals until today
\end{itemize}
The space heating demand is calculated based on the seasonal method for energy performance of buildings  \gls{EN} \gls{ISO} 13790 \cite{iso_standard} (see the supplementary material for further details). It is composed of the total heat losses (including losses by ventilation, thermal bridging and heat transfer) and the heat contributions (by solar radiation and internal heat gains). The heat savings $\Delta E_n$ at node $n$ are calculated by the ratio of space heat demand after and before the refurbishment (e.g. $\Delta E_n=0.6$ corresponds to 40\% space heat demand savings).  U-values, which measure how well a building element is insulated, and heated floor areas are used from the hotmaps project \cite{hotmaps} and the \gls{EU} buildings database \cite{EU_building}. Typical building topologies with corresponding standard values are taken from \gls{tabula} \cite{Episcope_tabula}. The ratio for additional insulation material between roof, floor and exterior wall is fixed determined by averaged monitored rates in \cite{IWU_lweights}. Thereby, they depend only on one common variable $\delta L$. The additional insulation material $\delta l_e$ [m] for a single  building element $e \in$ [roof, floor, wall] is then composed of the common variable $\delta L$ weighted by the  proportional expansion $f_e$.  Poorly insulated windows are replaced with double glazing in the moderate case and triple glazing in the ambitious case. The corresponding costs $c_{\text{retro}}$ are composed of fixed costs $c_{\text{fix}}$ and variable costs $c_{\text{var}}$  depending on the amount additional insulation material $\delta l_e$ for building element $e$ and costs for window replacement $c_{\text{window}}$
\begin{equation}
c_{\text{retro}}(\delta l) = \sum_{e} \left( c_{e, \text{fix}} + \delta l_e \cdot c_{e, \text{var}} \right) + c_{\text{window}}.
\end{equation}
 In the second step, the resulting cost per energy savings are stepwise linearised (see figure \ref{fig:energy_cost_curve}). The refurbishment measures are modelled with an additional supply technology at bus $n$ as described in equation \ref{eq:retro_gen} with above defined costs and a maximum heat output of $r_{n, max} = (1-\Delta E_n)$.  The moderate and ambitious retrofitting corresponds to averaged space heat demand of respectively 40\% and 20\% of the unrenovated state and average costs of 157 EUR/m$^2$ and 224 EUR/m$^2$. The costs for renovating the single components and their corresponding lifetime are taken from \cite{iwu}.
\subsection{Scenarios}
 In the first part (section \ref{sec:balance}), we focus on instruments that reduce the impact of peak heat demands. We consider the three compensating elements: building retrofitting, \gls{tes} and individual hybrid heat pumps with backup gas boilers. In order to analyse both the individual and the combined effects, all eight possible combinations of these three tools are presented. They range from a complete \textbf{\hyperlink{retro+}{flexible}}  scenario including building renovation, \gls{tes} and individual gas boilers to a \textbf{\hyperlink{rigid}{rigid}} one without any balancing elements (see table \ref{tab:scenarios}). In the second part (section \ref{sec:base}) we examine to what extent the peakedness of the heat profiles determines the technology mix as opposed to the over energy demand. In a thought experiment, the total heat demand is distributed evenly over the entire year instead of having a time-varying profile, in order to assess the impact of the profile shape versus the yearly heat demand.
\begin{table}[h]
	\resizebox{\columnwidth}{!}{
		\begin{tabular}{llllll}
			\cline{4-4}
			\rowcolor{gray!15}
			\textbf{scenario}
			 & \textbf{\begin{tabular}[c]{@{}l@{}} building \\ retrofitting \end{tabular}}
			  & \multicolumn{1}{l|}{\cellcolor{gray!15}
				\textbf{\begin{tabular}[c]{@{}l@{}}thermal energy \\ storage \\ (\gls{tes})\end{tabular}}}
			& \textbf{\begin{tabular}[c]{@{}l@{}} individual \\ gas \\ boilers\end{tabular}}
			 & \multicolumn{1}{l|}{\cellcolor{gray!15}
				\textbf{\begin{tabular}[c]{@{}l@{}}heat \\ demand \\ profile\end{tabular}}}
			 \\ \cline{4-4}
			 \hline
			\textbf{\hypertarget{retro+}{flexible}} &\cmark &\cmark &\cmark  &\cmark   \\
			\rowcolor{blue!15}
			\textbf{retro+igas} &\cmark &\xmark &\cmark &\cmark   \\
			\textbf{\hypertarget{retro+tes}{retro+tes}} &\cmark &\cmark &\xmark  &\cmark   \\
			\rowcolor{blue!15}
			\textbf{retro} &\cmark &\xmark &\xmark  &\cmark   \\
			\textbf{igas+tes} &\xmark &\cmark &\cmark  &\cmark   \\
			\rowcolor{blue!15}
			\textbf{igas} &\xmark &\xmark &\cmark  &\cmark   \\
			\textbf{\hypertarget{tes}{tes}}  &\xmark &\cmark &\xmark &\cmark   \\
			\rowcolor{blue!15}
			\textbf{\hypertarget{rigid}{rigid}} &\xmark &\xmark &\xmark  &\cmark   \\
			\arrayrulecolor{gray!70}\hline
			\textbf{flexible-evenload} &\cmark &\cmark &\cmark &\xmark   \\
			\rowcolor{black!30!green!15!white}
			\textbf{rigid-evenload} &\xmark &\xmark &\xmark &\xmark   \\
		\end{tabular}
	}
	\caption{Properties of the presented scenarios.}
	\label{tab:scenarios}
\end{table}
\section{Results}
\subsection{Balancing peak demands with building retrofitting, thermal energy storage (\gls{tes}) and individual gas boilers}\label{sec:balance}
In the following section effects on total system costs and technology composition of the three balancing instruments: retrofitting of the thermal envelope of buildings, thermal energy storage (\gls{tes}) and individual gas boilers as backup for heat pumps are compared. All eight possible scenario combinations of those three are presented. They range from a completely \textbf{\hyperlink{retro+}{flexible}} scenario including building renovation, \gls{tes} and individual gas boilers to a  \textbf{\hyperlink{rigid}{rigid}} scenario without any of the three elements. No further expansion of the electric transmission grid is considered. \\
\begin{figure}[!ht]
	\centering
	\includegraphics[width=1.0\linewidth]{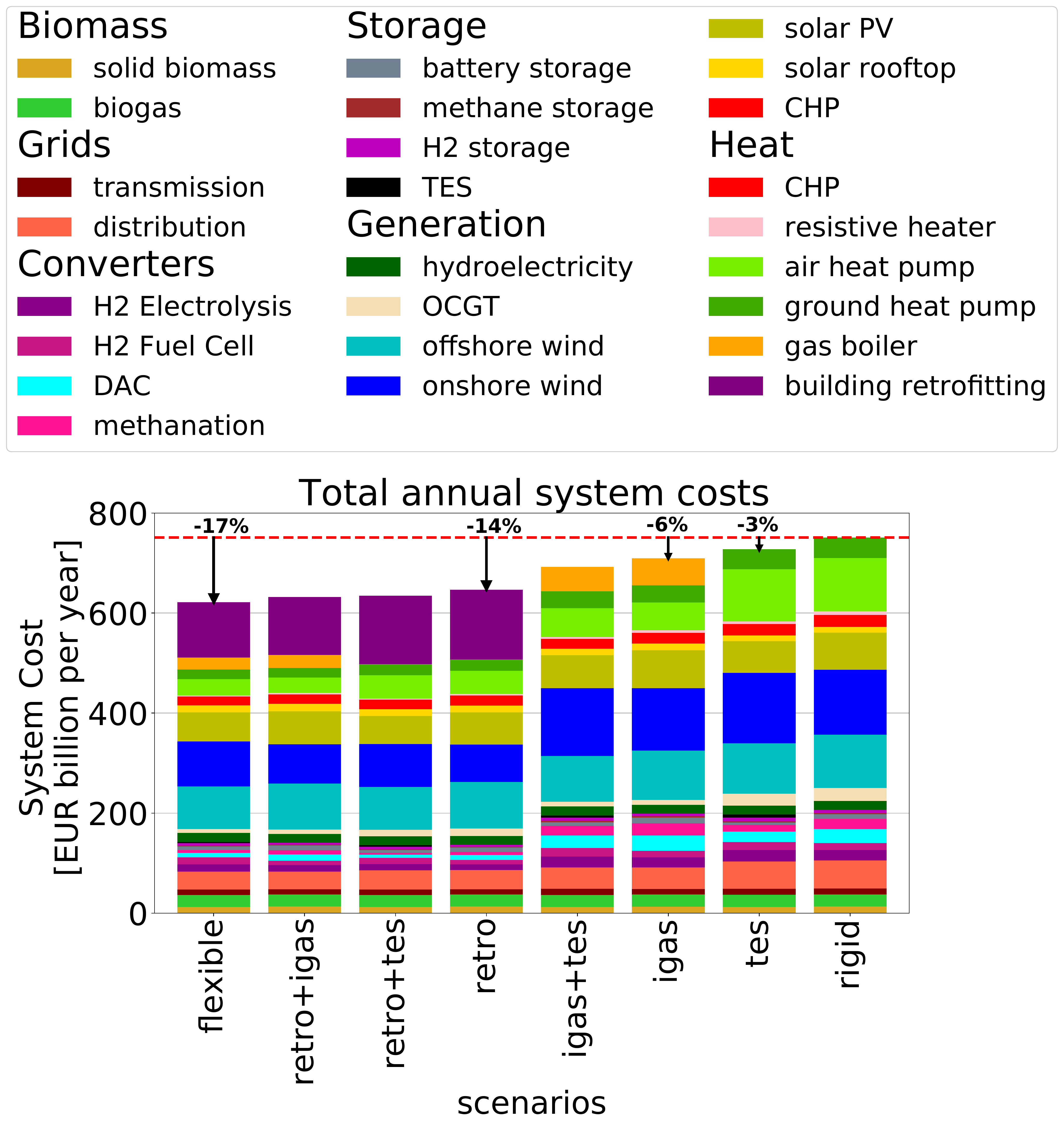}
	\caption{Total system costs for the 8 different combinations of the 3 balancing tools: building retrofitting, thermal energy storage (\gls{tes}) and individual gas boilers. Building retrofitting (\textit{4 left scenarios in the graph}) reduces overall expenses by up to 14\%. \gls{tes} and individual gas boilers, especially in the absence of building renovation, contribute to cost savings of up to 3\% and 6\% respectively. If building retrofitting is included, the availability of individual gas boilers has no major effect on total system costs (details in section \ref{sec:balance}).}
	\label{fig:balance_costs}
\end{figure}
\subsubsection{Costs}\label{sec:balance_cost}
 Overall costs decrease from 751 to 622 billion Euro per year as the number of available instruments for balancing peak demand increases (see figure \ref{fig:balance_costs}). Building efficiency improvements have the strongest leverage with up to 104 billion Euro savings per year. The availability of individual backup gas boilers has a more modest impact. In scenarios without building renovation, these provide a 6\% overall cost reduction. If retrofitting of the thermal envelope is in place, its availability barely affects the overall costs. In
this case, the gas distribution network to buildings can be removed and parts of the existing transmission network could be converted to the use of hydrogen for industrial processes or as back-up for electricity. \gls{tes} have the smallest effect on total system costs. Their presence leads to a maximum reduction of 3\%.
\subsubsection{Thermal supply}\label{sec:balance_thermal_supply}
\begin{figure*}[!ht]
	\centering
	\includegraphics[width=1.0\linewidth]{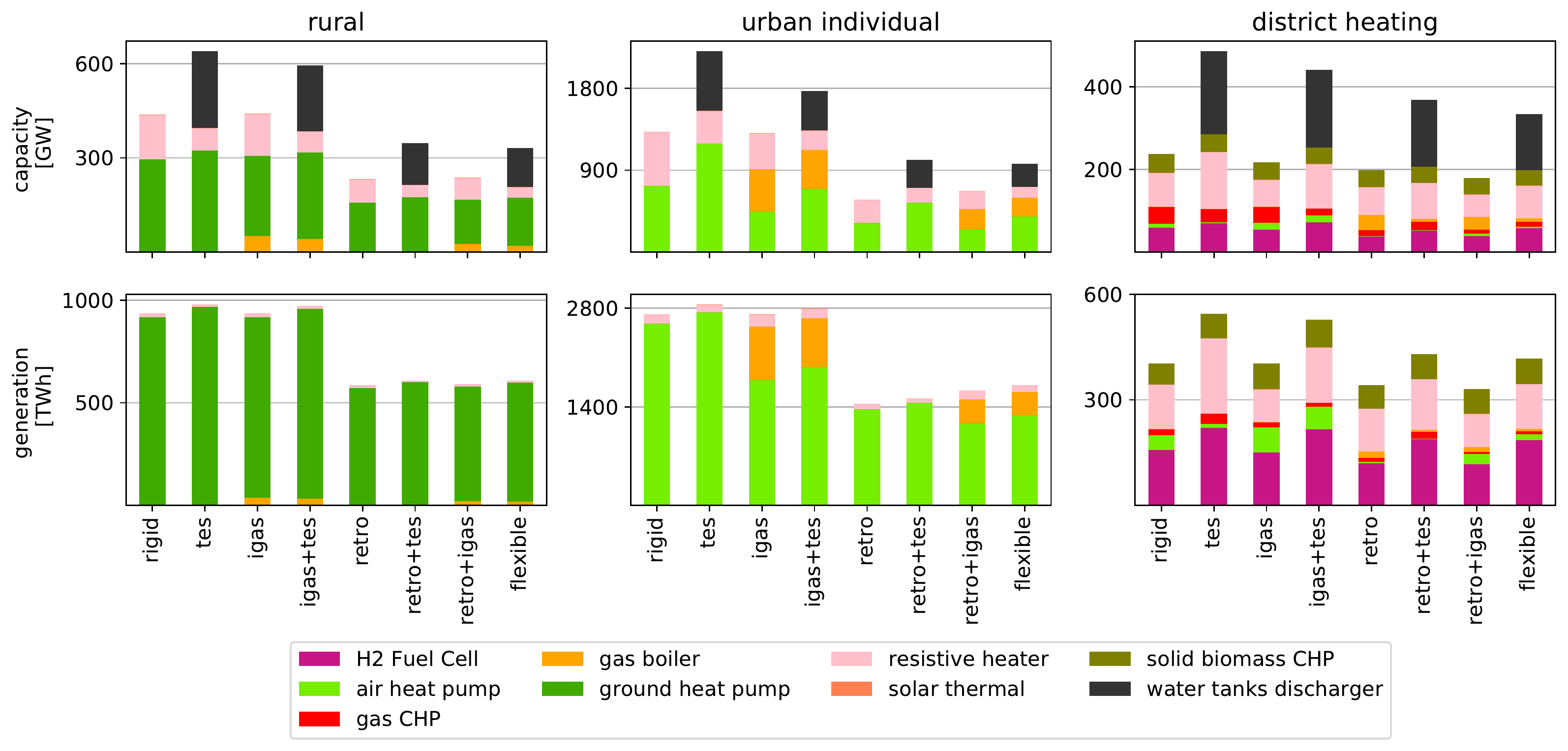}
	\caption{Installed peak capacities and thermal supply in rural areas and urban areas with individual heating and with district heating for 8 different scenarios ranging from a \textbf{rigid} one (far left) with no balancing instruments to a \textbf{flexible} scenario (far right) which includes building renovation, \gls{tes} and individual gas boilers (details in section \ref{sec:balance_thermal_supply}).}
	\label{fig:balance_thermal_supply}
\end{figure*}
\begin{figure*}[!ht]
	\centering
	\includegraphics[width=1\textwidth]{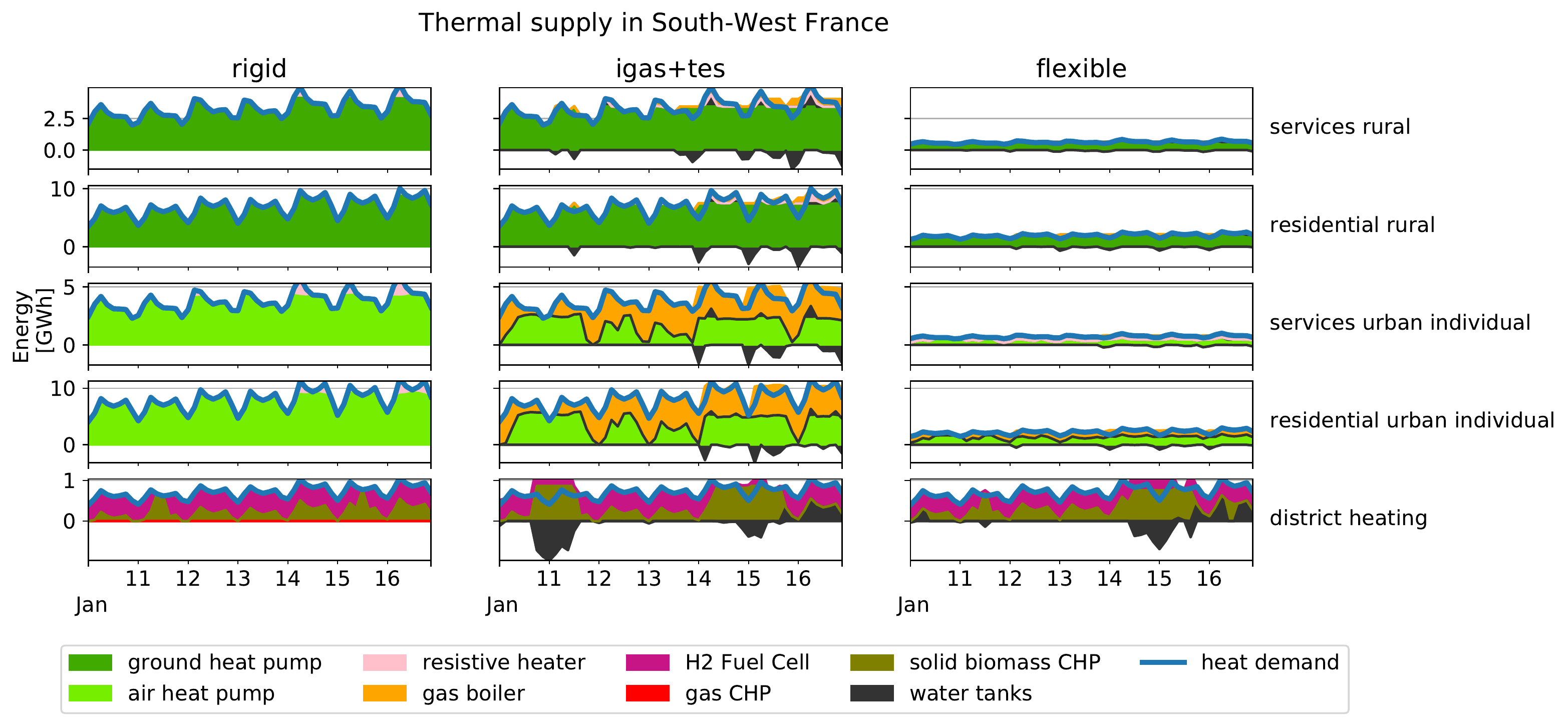}
	\caption{Example week of thermal supply at one node in South-West France for different heating systems (\textit{rows}) and scenarios (\textit{columns}). Peak heat demands are covered with backup gas boilers for individual heating in scenarios where these are available and no building retrofitting is in place (scenario \textbf{igas+tes}), especially in urban individual heated areas.  \gls{tes} is used in all scenarios (here shown  \textbf{igas+tes} and \textbf{flexible}) where these are in place to smooth peak demands. Strong heat demand savings are found for individual heating, but no renovation of the thermal envelope at this node in urban areas with district heating (details in section \ref{sec:balance_thermal_supply}).}
	\label{fig:balance_example_week}
\end{figure*}
\begin{figure}[!ht]
	\centering
	\includegraphics[width=1.0\linewidth]{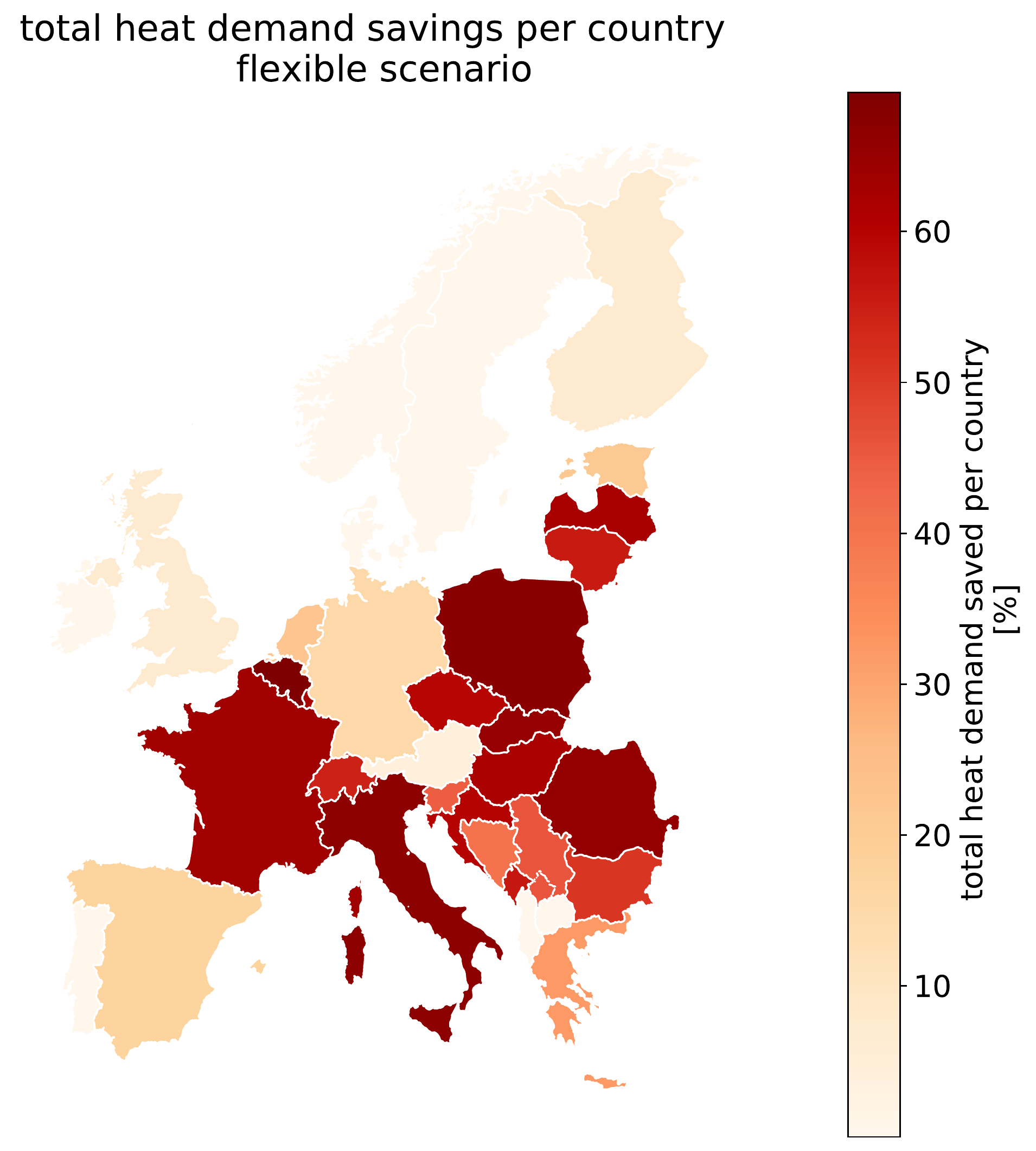}
	\caption{Total heat demand saved per country in the scenario \textbf{flexible} with all balancing instruments. Heat savings depend on the combination of both the country-specific costs for refurbishment and the costs for energy supply during the heating season (further details in section \ref{sec:balance_thermal_supply}).}
	\label{fig:heat_saved_ct}
\end{figure}
\begin{figure}[!ht]
	\centering
	\includegraphics[width=1.0\linewidth]{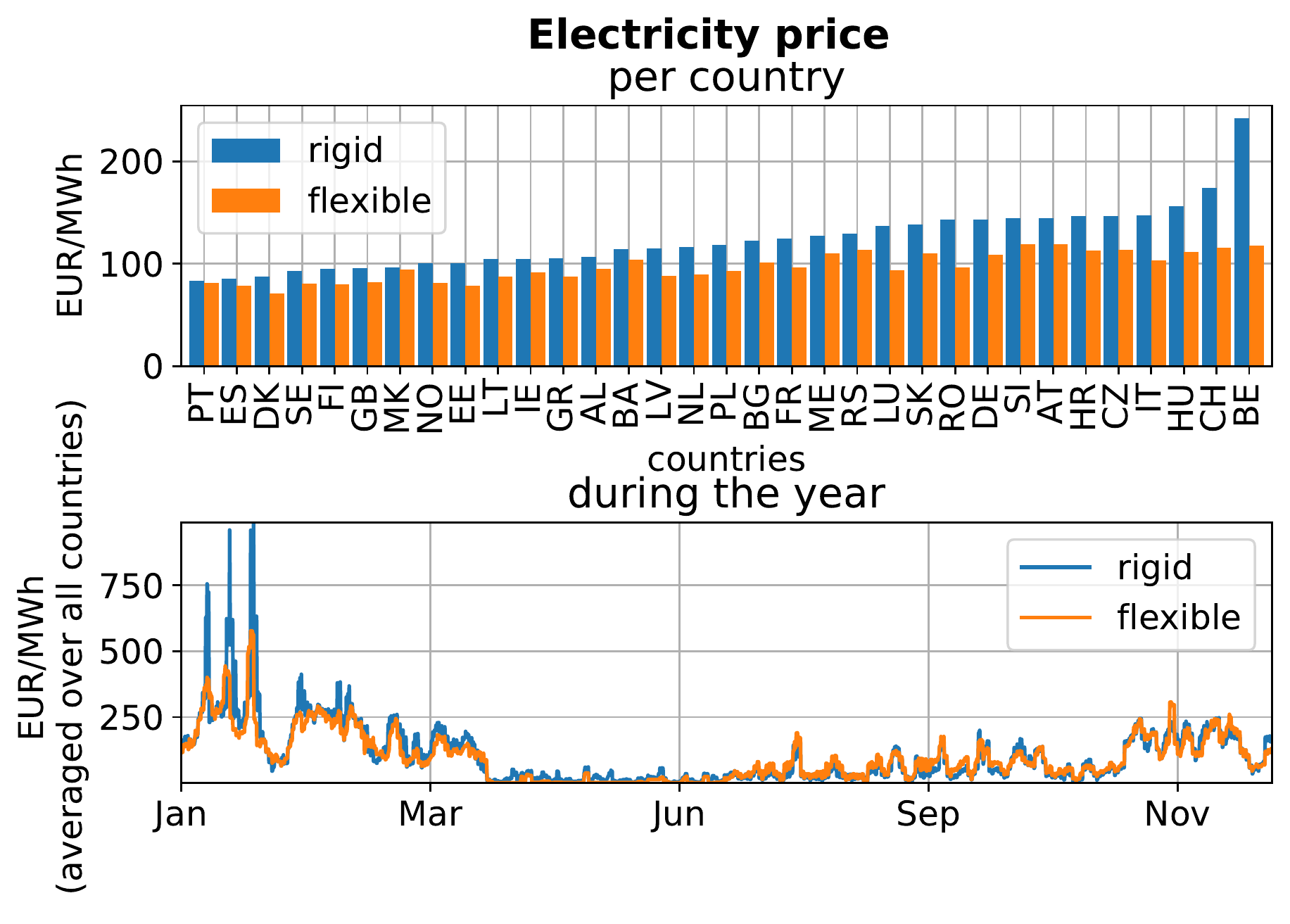}
	\caption{Marginal prices for electricity obtained from the \gls{KKT} multipliers weighted by load for the different countries in the \textbf{rigid} and \textbf{flexible} scenario. Retrofitting strength depends on the prices for electricity since large part of the heating sector is electrified. The electrification of the heating sector is also reflected in the marginal prices of electricity during the year which are higher in winter months (further details in section \ref{sec:balance_thermal_supply}).}
	\label{fig:balance_marginal_prices}
\end{figure}

Depending on the scenario, between 2113 TWh in scenario \textbf{\hyperlink{retro+tes}{flexible}} and 4379 TWh  heat in scenario \textbf{\hyperlink{tes}{tes}} are supplied to buildings. The scenario \textbf{\hyperlink{tes}{tes}} has a higher heat demand than \textbf{\hyperlink{retro+tes}{rigid}} because of losses in storage. The largest part, 75-95\%, is generated from electricity (heat pumps, resistive heater), 4-24\% from gas (gas \gls{chp}, gas boiler, fuel cell heat) and 1-3\% from biomass (solid biomass \gls{chp}). Solar thermal generators are included, but are hardly used (maximum 4 TWh). The balancing instruments reduce energy demand by up to 51\% and peak capacities by up to 55\%. Of the three balancing options considered, efficiency improvements in the building sector have the greatest impact on peak capacity reduction (see figure \ref{fig:balance_thermal_supply}). In sum, between 44-51\% of the space heat demand (36-42\% of total heat demand including hot water) are saved. \gls{tes} increases overall generation, but decreases peak capacities of gas boilers. If individual gas boilers are missing, their supply is mainly replaced by heat pumps and at peak demands by resistive heater (see figure \ref{fig:balance_example_week}). \\
\\In areas with individual heating, the major part of thermal supply is provided by heat pumps while gas boilers and resistive heaters are used as back-up capacity for peak demand. Thermal supply is most expensive in urban individual heated areas. Marginal prices based on the \gls{KKT} multiplier $\lambda$ from equation (\ref{eq:KKT}) vary between 106-138\mwth, compared to rural ones with 90-108\mwth. This difference between rural and urban areas is caused by  the use of air-sourced instead of ground-sourced heat pumps in urban areas due to land restrictions.  Compared to ground-sourced heat pumps, their efficiency depends on the outside and not on the more stable ground temperature and thus correlates more strongly with the heating demand. Especially on cold days the efficiency is lower than that of ground-sourced heat pumps. This leads to a higher contribution of gas boilers to individual heat supply in urban than in rural areas. Retrofitting of the thermal envelope is particularly advantageous due to the high costs for heat supply. \gls{tes} are used to smooth daily peaks. Their maximum capacity is reduced by up to 58\% if building retrofitting is in place.\\
\\ In areas with district heating the technology mix is more balanced. Heat is mainly supplied by fuel cells, resistive heaters and \gls{chp}s (solid biomass and gas). In scenarios without building retrofitting or \gls{tes} air heat pumps also cover up to 10\% of the supply. There is less building renovation with district heating compared to decentralised systems. This is caused by the possibility to install technologies on a large scale instead of individual solutions which results in lower marginal prices of on average 41\mwth. Therefore the reduction in generation is less pronounced. The largest fraction of \gls{tes} is installed in areas with district heating (85-95\% of the total energy capacity) since there they can be used as seasonal storage. The lack of individual gas boilers also indirectly influences the technology mix in the district heating network. Their absence leads to higher electricity prices, marginal prices increase for scenarios without individual gas boilers by on average 4\mwh (with retrofitting) and 13\mwh (without retrofitting) and reduce the use of heat pumps. \\
\\Geographically, the amount of saved space heat differs strongly between countries (see figure \ref{fig:heat_saved_ct}). The strength of building renovation depends on the interplay between the costs of refurbishment and those for energy supply during the heating season. The cost of refurbishment is based on the country-specific building stock characteristics and expenditure on labour and taxes. They are for example low in Romania, Slovakia and Poland, where more than 60\% of the heat demand is saved. Electricity supply prices are higher in winter months and are especially high on weather events with cold dark wind lulls (see figure \ref{fig:balance_marginal_prices}). In countries with high electricity costs like Belgium, Switzerland and Hungary heat savings of more than 54\% are attractive despite high renovation costs. As a large part of the heat sector is electrified, the rate of building renovation also depends on the technology mix of electricity generation in each country. The seasonal generation from wind is better suited to peak heat demand in winter months than solar \gls{PV}. Countries with a large share of wind generation, such as Great Britain, Denmark or Portugal, have cheaper electricity in winter and therefore a lower renovation as a result.
\subsubsection{Electricity generation and storage}
Electricity is generated mainly from renewable energies (solar, wind and hydro power), and to a minor extent of 3-8\% from dispatchable power plants using renewable gas (\gls{chp}, \gls{ocgt}). Wind and solar contribute the largest share of electricity generation with 90-95\%. While renovations in buildings and individual gas boilers reduce the total generation from 6725 TWh down to 4982 TWh, \gls{tes} increases the electric supply due to losses in storage. In scenarios including retrofitting, peak capacity in the electricity distribution network is reduced by 30\%.\\
\\The more balancing instruments are in place, the less installed wind and solar power is needed and capacity decreases by up to 26\%. Onshore wind declines particularly strongly in scenarios including building retrofitting with a decrease of up to 39\% from the \textbf{\hyperlink{rigid}{rigid}} to the \textbf{\hyperlink{retro+}{flexible}} scenario, while solar rooftop is increasing  with a plus in installed capacity of up to 29\%. The particularly sharp decline in onshore wind and growth in rooftop PV is caused by the seasonality of both demand and generation. The heat demand is high in winter because of increased space heating needs, while during this period the feed-in of solar is lower and the potential of onshore wind is higher. Supply of solar is therefore more profitable, if demand peaks in winter are reduced. This reduction of onshore wind capacity of up to 640 GW through building renovation, is not only more cost-efficient but also avoids problems of social acceptance of further onshore wind farms. In scenarios including building renovation, up to 30\% less electricity is fed back into the distribution grid through \gls{v2g}. In the presence of \gls{tes}, onshore wind capacities increase. The variability of onshore wind generation can be better smoothed out by pre-heating the buildings at times of high wind feed-in. The availability of \gls{tes} further reduces peak capacities of battery storage by up to 56\%.
\subsubsection{Hydrogen and methane}\label{sec:balance_converters}
\begin{figure}[!ht]
	\centering
	\includegraphics[width=1.0\linewidth]{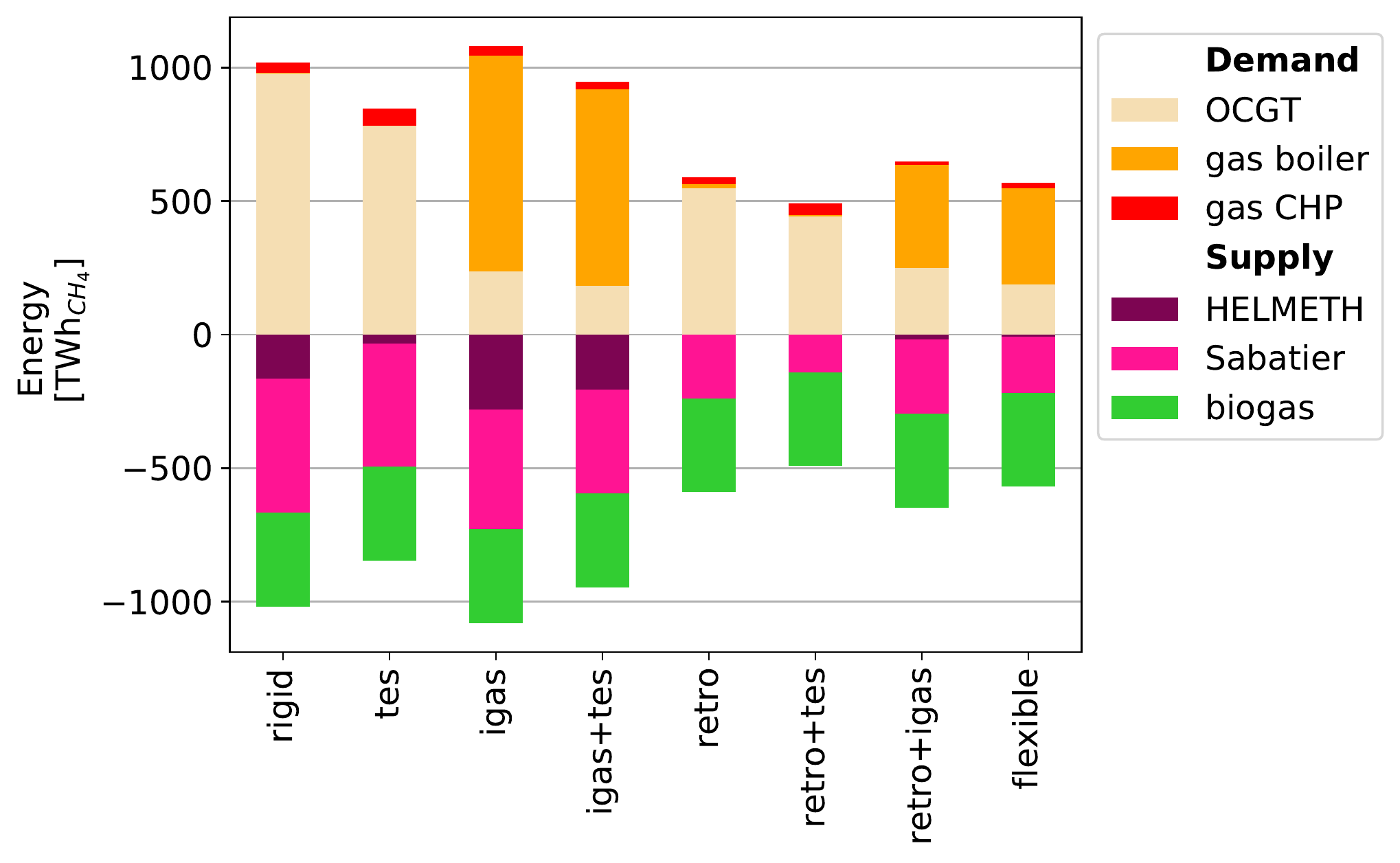}
	\caption{Balance of the gas demand (for \gls{ocgt}, gas boiler and gas CHP) supplied by methanation processes (Sabatier, \gls{helmeth}) or upgraded biogas. Gas demand is almost halved by building retrofitting (see section \ref{sec:balance_converters}).}
	\label{fig:gas_balance}
\end{figure}

In scenarios including renovation of the thermal envelope, the need for back-up hydrogen and methane is significantly reduced. The balancing elements reduce hydrogen production from 1067 to 498 TWh$_{\text{H}_2}$, synthetic methane from 728 to 95 TW$_{\text{CH}_4}$ and the conversion of hydrogen to electricity from 245 to 120 TWh$_{\text{elec}}$.  \\
\\The use of hydrogen depends on the installed onshore wind power and the space heat demand in buildings. Higher wind capacities result in a larger surplus of electrical energy at favourable weather conditions with low demand, which can be used in the electrolysis for conversion into hydrogen. Therefore the capacities of electrolysis and fuel cells are highest in the presence of \gls{tes} and in the absence of building renovation with 453 GW and 161 GW respectively.\\
\\ The total methane demand for gas boilers, gas \gls{chp} and \gls{ocgt} decreases most strongly in scenarios including building renovations, by an average of 48\%, \gls{tes} enables a reduction of 17\% and the absence of individual gas boilers of 7\%. The minor reduction in gas demand in the absence of individual gas boilers can be explained by the increased use of \gls{ocgt} (see figure \ref{fig:gas_balance}). \\
\\In all scenarios, the full potential of upgraded biogas, 352 TWh, is exploited. The rest is met by synthetic gas from a methanation process. The Sabatier process is used in all scenarios but the installed capacities are decreasing by up to 82\% in scenarios including retrofitting. The \gls{helmeth} process, which combines energy-efficiently the two steps of first converting electricity into hydrogen and second hydrogen into methane, is only used in scenarios without building renovation with maximum installed capacity of 56 GW. The lower use of \gls{helmeth} is caused by higher investment costs compared to the Sabatier process therefore the \gls{helmeth} process needs more full load hours to be built.\\
%
%
\subsection{A thought experiment - peak vs. constant heat load\textit{}}\label{sec:base}
A thought experiment is discussed in this section which serves to illustrate the effects of peak demands on overall costs and system composition, though is not intended to be a realistic scenario. Instead of using time-varying load profiles, annual thermal demand is distributed evenly over the entire year. Thus the total amount of heating demand over the year is the same, but there is only a constant heat profile without the winter peak. This allows us to distinguish between measures that are implemented to reduce overall energy demand, and those that purely serve to reduce the peak. \\
\begin{figure}[!ht]
	\centering
	\includegraphics[width=1.0\linewidth]{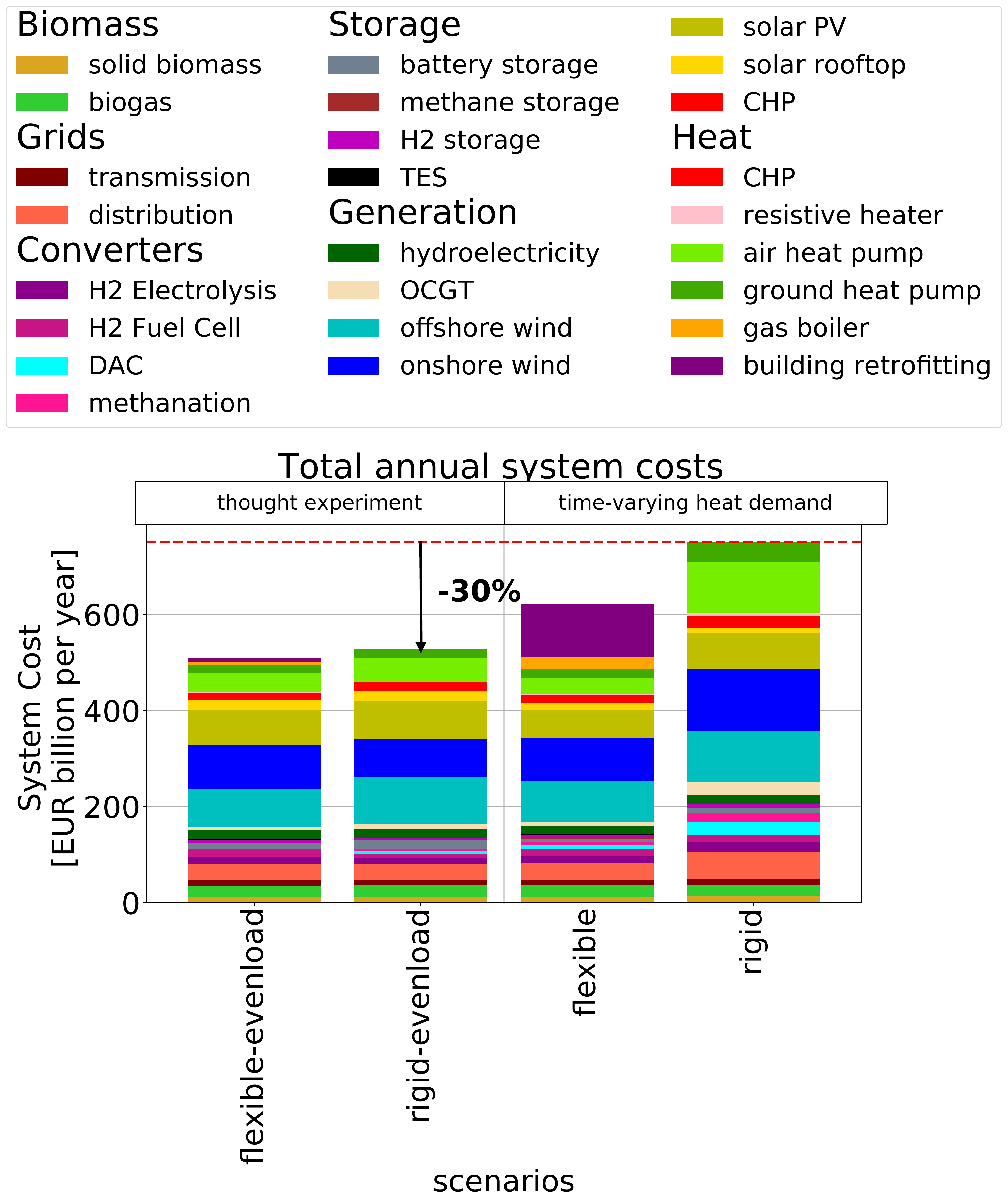}
	\caption{Total system costs of scenarios with an assumed flat heat demand (\textit{two left}) and with heat demand profile (\textit{two right}). Costs with a constant load during the year are up to 30\% lower (see section \ref{sec:base_costs}).}
	\label{fig:costs_baseload}
\end{figure}
%
\label{sec:base_costs}
\\ Total costs decrease between 18-30\% compared to scenarios with load profile  (see figure \ref{fig:costs_baseload}). This reveals that about a quarter of the entire costs of the three sectors heating, electricity and transport, are generated by heat peak demands. With a constant heat demand distributed throughout the year, there is barely any building renovation, \gls{tes} energy capacities are decreasing by up to 75\% and no gas boilers are used in rural areas. Scenarios which include varying balancing tools share similar costs and system compositions in the  case of a constant load. It underlines that the considered instruments (building efficiency, thermal energy storage and individual gas boiler) are primarily used to balance thermal peak demands rather than reducing energy demand. In case of a constant heat demand, their usage is strongly reduced and their availability does not lead to a significant difference of system costs and design.\\
\begin{figure}[!ht]
	\centering
	\includegraphics[width=1\linewidth]{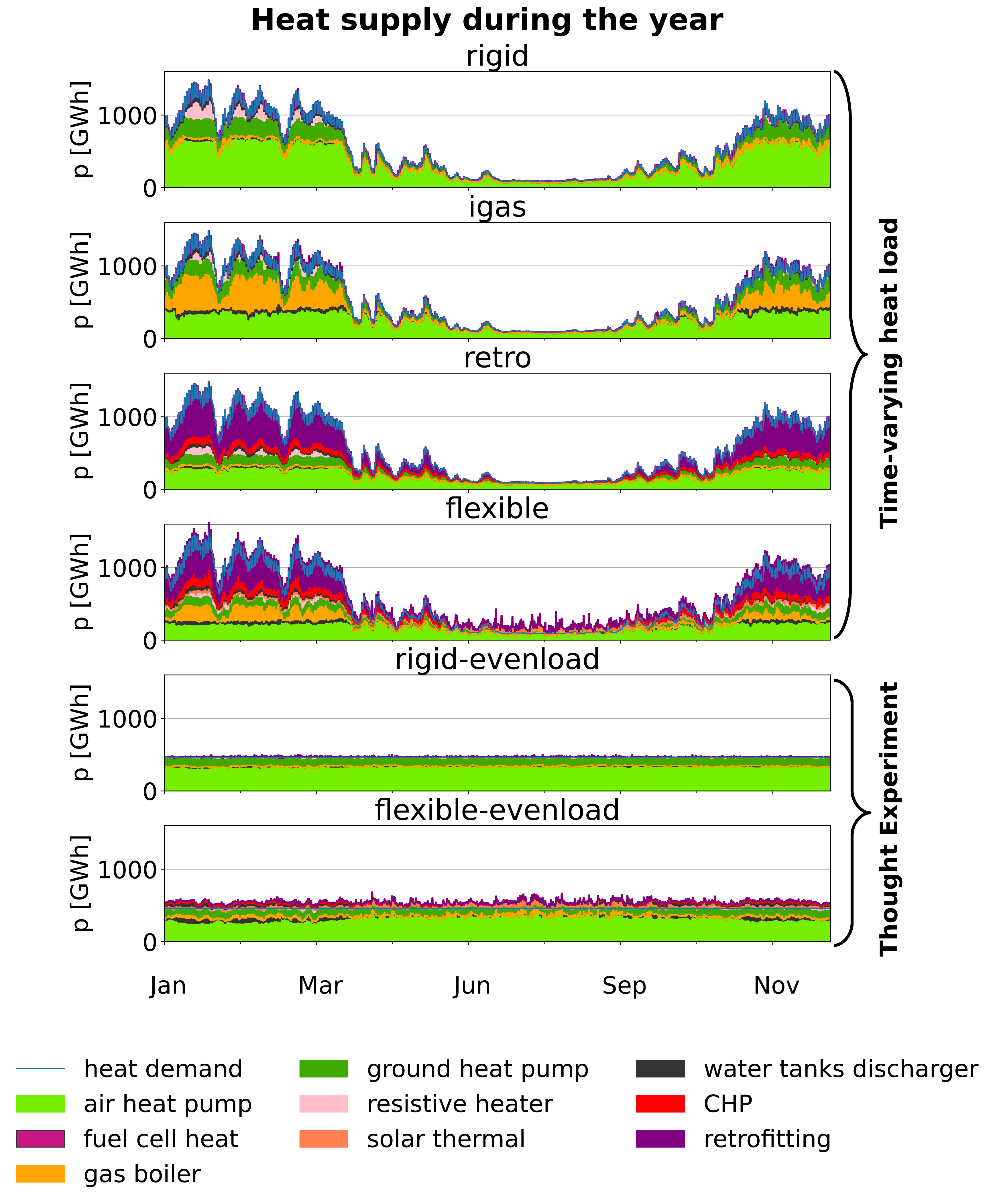}
	\label{fig:heating_p}
	\caption{Thermal supply in Europe during the year for scenarios with time-varying heat demand and for the two thought experiment scenarios with a heat demand as a constant load. The saved space heat demand (\textit{purple}) is decreasing by 6-8\% space heat savings in the case of a constant heat demand. This shows that building retrofitting is mostly important to smooth demand peaks and not to reduce overall energy demand (further details in section \ref{sec:base_thermal_supply}).}
\end{figure}
\label{sec:base_thermal_supply}
\\Thermal supply of a constant load is covered to a greater extent by heat pumps. In turn the share of heating supplied by gas boilers, resistive heaters and \gls{chp}s declines by 7\%, 5\% and 1\% respectively. In densely populated areas the supply of heat pumps increases by 14-17\%, in rural areas they cover all demand. As the heat sector is fully electrified in rural areas, a gas grid in these regions is redundant. The amount of heat provided by fuel cells lies at around 6\% for all scenarios and depends on the proportion of installed onshore wind power rather than on the heat load profile. Building retrofitting plays a subordinate role in our thought experiment. Depending on the scenario, 6-8\% of the space heat demand are saved (compared to 44-51\% with a thermal profile), mainly in Eastern European countries. Renovation of the building stock is thus primarily required to manage the peak load and not to reduce total energy demand. \gls{tes} drops to about 7.8 TWh in all scenarios, which corresponds to a reduction of 56-75\%. This indicates that a large part of the storage capacity is used to balance the fluctuating demand and not to compensate the variable renewable generation. \\
\\The composition of installed solar and wind capacities is similar in all scenarios and adds up to about 3 TW without transmission expansion. Solar, especially rooftop \gls{PV} systems, is favoured in case of a base load with a capacity increase of 14-24\% (rooftop \gls{PV} of 30-47 \%) compared to the scenarios with a heat load profile. The upswing of solar is caused by the absence of any seasonality in heat demand. Typically, thermal load is high in winter when the potential for solar system is low. \gls{PV} becomes more advantageous if heat demand is reduced in winter times (as in scenarios including retrofitting and \gls{tes}) or even remains constant over the entire year (as in this thought experiment). Offshore wind capacities stay constant and are independent of the heat load profile. Onshore wind capacities do not follow such a clear trend. Installed onshore capacity increases in scenarios including building retrofitting by up to 10\%, while in the others it decreases by up to 66\%. This indicates that with a steady load, onshore wind farms are more competitive compared to efficiency improvements in the building sector. Battery and hydrogen storage depend on the installed capacity of respectively solar and wind power. As the proportion of solar power is higher in all cases, battery capacity rises by 40-45\%. \\
%

\FloatBarrier
\section{Discussion}
\subsection{Comparison to literature}
Consistent with other studies \cite{heat_roadmap3, staffell2018}, our results show an increase in electricity peak demand when buildings are decarbonised. In particular, cold, windless days pose a challenge and rise total system costs, as Staffell et al. \cite{staffell2018} have shown for Great Britain. Our results support  Staffell's statement that a high temporal resolution of demand and supply is necessary in order to capture the challenge of these heat peak demands. This study further expands our knowledge by quantifying the costs and effects of these challenging heat peaks in a European system. \\
\\Malley et al. \cite{malley} investigated for Ireland, as well as Ashfaq et al. \cite{ashfaq2018} for the Danish city Aarhus, that \gls{tes} allows a higher wind-feed in. The same effect is evident in our European model. In agreement with Hedegard \cite{hedegaard2013}, our results indicate that additional thermal storage allows the flexible operation of heat pumps and thus leads to a cost decrease as well as a reduction of the necessary peak capacities. However, with cost savings of a maximum of 3\%, these advantages are significantly lower in our results than the findings from Hedegaard. This could be caused by the additional flexibility of coupling different sectors and the possibility to balance challenging weather events spatially. In addition, we capture the competition of several balancing instruments. \\
\\ In this study peak capacities are reduced significantly more by hybrid heat pumps with backup gas boilers or building efficiency improvements than by \gls{tes}. Our results on a European level agree with findings of Eggimann et al. \cite{eggimann2020} for the UK. In this work, four scenarios were examined with different levels of electrification and fuel switching. It is shown consistent with our findings that the combination of building renovation and the use of gas-to-heat can reduce heat peak demand. Chen et al. \cite{chen2021} find for Northern Europe the use of hybrid heat pump with backup gas boilers for individual heating during peak demands advantageous if carbon-neutral gas is available. In contrast to Eggimann et al. and Chen et al., in this study the increase in building efficiency is co-optimised. We can thus also determine the cost-optimal balance between hybrid heat pumps and investments in building renovation. \\
\\  Building renovation is found in most studies to be a key element to reduce overall costs. In agreement with our results, on a European level, heat savings of 30-50\% are found in \cite{heat_roadmap3, Broin2013, jrc_times2017}. Consistent with the presented results in our study, Kotzur et al. \cite{kotzur2020} find in scenarios without gas supply only a minor increase of total system costs of 3.65\%.

\subsection{General discussion}
There are numerous other benefits of building renovation and abandoning individual hybrid heat pumps with backup gas boilers which are not considered in our study but have an impact on the energy system composition. Renovated buildings lead for example to a higher thermal comfort, noise reduction, reduced fuel poverty, increased energy security due to lower consumption or additional employment \cite{retro_jobs}, which also can be seen as a post \gls{covid} stimulus \cite{IEA_covidstimulus}. Renovation further improves the health of inhabitants as it leads for example to lower mortality rates in winter, less cardiovascular and respiratory diseases and less depression \cite{health_retro, health_retro2,health_retro_nz, health_retro_england, health_retro_england2}. These are all relevant aspects which could lead to higher rates of renovation than those presented in our results. \\
\\ There are equally other reasons for the abandonment of individual hybrid heat pumps with backup gas boilers. First, some of the methane is released during storage, transmission or burning in the gas boilers \cite{gas_network_uk, gas_transmission_leak}. Methane is an aggressive greenhouse gas and high leakage rates would make the use of gas boilers unreasonable from a climate protecting perspective. Second, the less consumers using the gas network, costs for the remaining are increasing, what could increase the number of households in fuel poverty. Third, Korzacanin et al.  \cite{pumps_vs_boiler} show that climate change makes heat pumps more attractive compared to gas boilers. On the one hand they do have a higher efficiency with higher ambient temperatures. On the other hand, reduced space heat demand increases the share of the hot water demand, which is constant over the year. For a constant demand heat pumps are more suitable. Finally, it simplifies the connection of new buildings, as they only need to be connected to electricity and water.

\subsection{Sensitivity analysis and limitations}

See the Supplementary Information for a sensitivity analysis concerning different weather years, costs, electricity grid expansion, district heating share and sink temperature of heat pumps, as well as a list of limitations of the model.

\section{Conclusion}
Decarbonisation of the heat sector is a challenging task, as electrification makes both demand and supply dependent on weather conditions. In this paper we discuss how to deal with the impacts of strong seasonal space heating demand peaks in the most cost-effective way. By modelling with hourly resolution for the whole of Europe, we capture the critical events of simultaneous high heat demand an low renewable feed-in. Unlike other models that decouple supply from efficiency considerations, we have co-optimised demand and supply side measures to explore the competition between them  in a sector-coupled European model with net-zero CO2 emissions. All of the considered methods to mitigate heating peaks, consisting of building renovation, thermal energy storage (\gls{tes}) and individual heat pumps with backup gas boilers, lead to a reduction of the total system costs by up to 17\%. Their usage depends on the seasonal interaction between generation and heat demand.\\
\\ Building renovations, which result in 44-51\% space heat demand savings, show the strongest effect  on the reduction of total costs with 14\% cost savings. Individual gas boilers that are used as back-up for heat pumps, can be completely removed in rural areas with a minor cost increase of only 1-2\% if the buildings are renovated.  \gls{tes} on the one hand smooths out peak demands and on the other hand enables a higher feed-in of variable generation. Overall thermal supply with \gls{tes} is larger because of storage losses, but total system costs are reduced by at most 3\%. \\
\\ Through the thought experiment in which a constant heat profile is assumed instead of a varying one, it is shown that up to 30\% of the total costs are generated by heat peak demands. The system optimum with a constant heat profile has only an 8\% saving of heat demand compared to the correct peaked profile. This means building renovation is mainly needed to manage those peaks and not to reduce the total energy demand. Up to 75\% of the energy capacity of \gls{tes} is used to mitigate peak demands while the remaining part balances the variable generation. \\
\\ In conclusion, demand peaks of space heating in a highly renewable scenario may seem like an insurmountable challenge, but in this paper we have shown that several strategies can be combined to mitigate their cost impacts.
\section{Data availability and code availability}
The source code and all the input data zenodo/github, documentation of PyPSA \cite{pypsa_doc}.
\section{Acknowledgement}
The authors thank Marta Victoria P\'erez, Alexandru Nichersu, Johannes Hampp, Fabian Neumann, Martha Frysztacki, Fabian Hofmann and Amin Gazafroudi for helpful discussions
and suggestions. T.B. and E.Z. acknowledge funding from the Helmholtz Association under grant no. VH-NG-1352. The responsibility for
the contents lies with the authors.
\section{References}
\bibliography{bib_retro}
\clearpage
\newpage\null\thispagestyle{empty}
\printglossary[type=\acronymtype]
\clearpage
\newpage\null\thispagestyle{empty}

\end{document}